\documentclass[twocolumn,floatfix,prb,superscriptaddress,longbibliography]{revtex4-1}
\usepackage{graphicx}% Include figure files
\usepackage{dcolumn} % Align table columns on decimal point
\usepackage{bm}
\usepackage{epsfig}
\usepackage{longtable}
\begin{document} 

\title{Averievite: a copper oxide kagome antiferromagnet}

\author{A. S. Botana}
\affiliation{Materials Science Division, Argonne National Laboratory, Argonne, Illinois 60439, USA}
\author{H. Zheng}
\affiliation{Materials Science Division, Argonne National Laboratory, Argonne, Illinois 60439, USA}
\author{S. H. Lapidus}
\affiliation{X-ray Science Division, Advanced Photon Source, Argonne National
Laboratory, Argonne, Illinois 60439, USA}
\author{J. F. Mitchell}
\affiliation{Materials Science Division, Argonne National Laboratory, Argonne, Illinois 60439, USA}
\author{M. R. Norman}
\email{norman@anl.gov}
\affiliation{Materials Science Division, Argonne National Laboratory, Argonne, Illinois 60439, USA}
\date{\today}

\begin{abstract}

Averievite, Cu$_5$V$_2$O$_{10}$(CsCl), is an oxide mineral composed of Cu$^{2+}$ kagome layers
sandwiched by Cu$^{2+}$-V$^{5+}$ honeycomb layers.
We have synthesized this oxide and
investigated its properties from \textit{ab initio} calculations along with susceptibility and specific heat measurements.
The data indicate a Curie-Weiss temperature of 185 K as well as long-range magnetic order at 24 K due to the significant interlayer coupling from the honeycomb copper ions.
This order is suppressed by substituting copper by isoelectronic zinc, suggesting that Zn-substituted averievite is a promising spin liquid candidate. 
A further proposed substitution that replaces V$^{5+}$ by Ti$^{4+}$ not only dopes the material, but is predicted to give rise to a two-dimensional electronic structure featuring Dirac crossings.  As such, averievite is an attractive platform for S=1/2 kagome physics with the potential for realizing novel electronic states. 

\end{abstract}

\maketitle

\section{Introduction}

Condensed matter physics and mineralogy have traditionally been separate spheres of endeavor. However, their intersection can lead to interesting new science, especially in the context of novel magnetism and spin liquid behavior.\cite{review_herb} A recent example is herbertsmithite, a copper hydroxychloride mineral with Cu$^{2+}$ (S=1/2) kagome layers separated by nonmagnetic Zn ions.\cite{herb1}  Despite its large Curie-Weiss temperature, herbertsmithite shows no evidence for long range magnetic order, an indication of quantum spin liquid (QSL) behavior.\cite{review_herb,herb1,balents1,balents2,zhou} Furthermore,
inelastic neutron scattering exhibits a broad spin continuum, consistent with fractionalized excitations.\cite{herb_exc}  Given the interest in doped QSLs,\cite{anderson2} herbertsmithite is an obvious target for chemical doping studies.   Doping with Ga on the Zn site was explored theoretically by Mazin \textit{et al.}~and is predicted to produce several novel states, including $f$-wave superconductivity, and at 1/3 electron doping, a correlated Dirac metal.\cite{mazin}  However, doping is not easy to achieve since the material is a hydroxide, with decomposition being the usual result.\cite{nytko} Lithium intercalation has been reported, but the material remains insulating,\cite{li_doping} perhaps due to polaron formation given the large Mott gap of the stoichiometric phase.\cite{mott}  Therefore, it would be desirable to find other
QSL candidates based on an S=1/2 kagome lattice that do not have these issues.

\begin{figure}
\includegraphics[width=\columnwidth,draft=false]{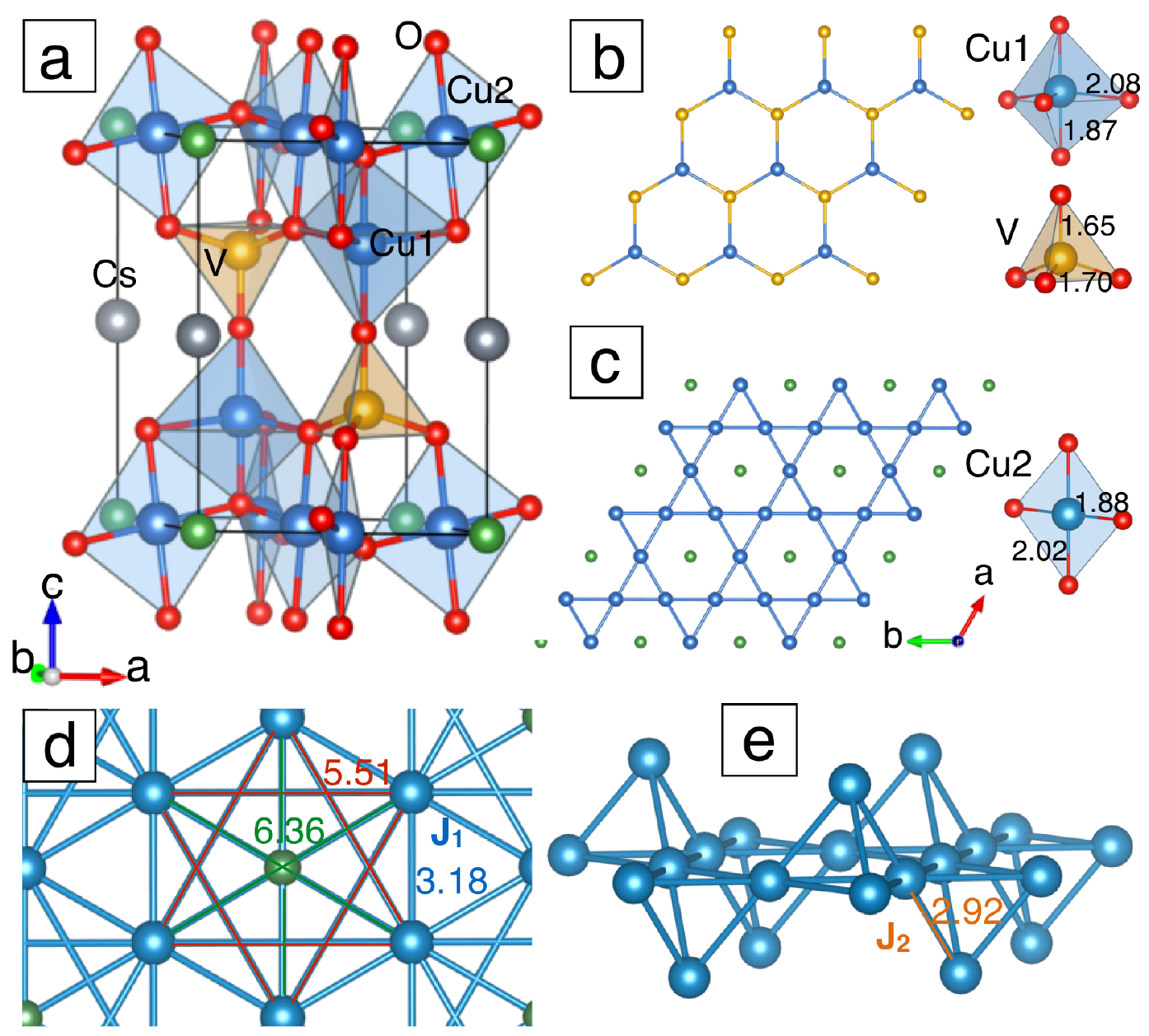}
\caption{(a) Structure of averievite, composed of CuVO$_3$ honeycomb layers (b) that sandwich Cu$_3$O$_2$Cl kagome layers (c) with different local environments for the Cu atoms (Cu-O and V-O distances are shown in \AA). Each of these three-layer blocks are separated by CsO$_2$ layers.  Cs atoms are in gray, Cu atoms in blue, O atoms in red, V atoms in yellow, and Cl atoms in green.  (d) Cu-Cu distances in the kagome plane (in \AA). (e) Cu lattice involving honeycomb and kagome layers, showing pyrochlore slabs reminiscent of clinoatacamite. The interlayer Cu-Cu distance is shown in \AA, and $J_1$ and $J_2$ denote magnetic couplings.}
\label{fig1}
\end{figure}

By searching the mineralogical literature, we have identified averievite, an oxide mineral that contains S=1/2 kagome planes that was found as a product of post-eruption volcanic activity.\cite{averievite}  The synthetic route to averievite has been reported in unpublished work by Queen,\cite{queen} where only structural data were shown.  Represented by the formula Cu$_5$O$_2$(VO$_4$)$_2$(MX)$_n$, where M is an alkali metal and X a halide, it has been grown with n=1 and MX=RbCl, CsCl and CsBr. Structurally, the material consists of Cu$_3$O$_2$Cl kagome layers in which each Cu bonds with four O atoms in a square planar coordination. The kagome planes are sandwiched by two CuVO$_3$ honeycomb layers composed of CuO$_5$ trigonal bipyramids and VO$_4$ tetrahedra (Fig.~\ref{fig1}).  The copper layers form a pyrochlore slab, similar to clinoatacamite, the parent phase of herbertsmithite,\cite{clino} but unlike it, a connected 3D network is not formed.  Rather, the trilayers in averievite are separated by CsO$_2$ layers, potentially leading to more two dimensional-like behavior.

\begin{figure}
\includegraphics[width=0.9\columnwidth,draft=false]{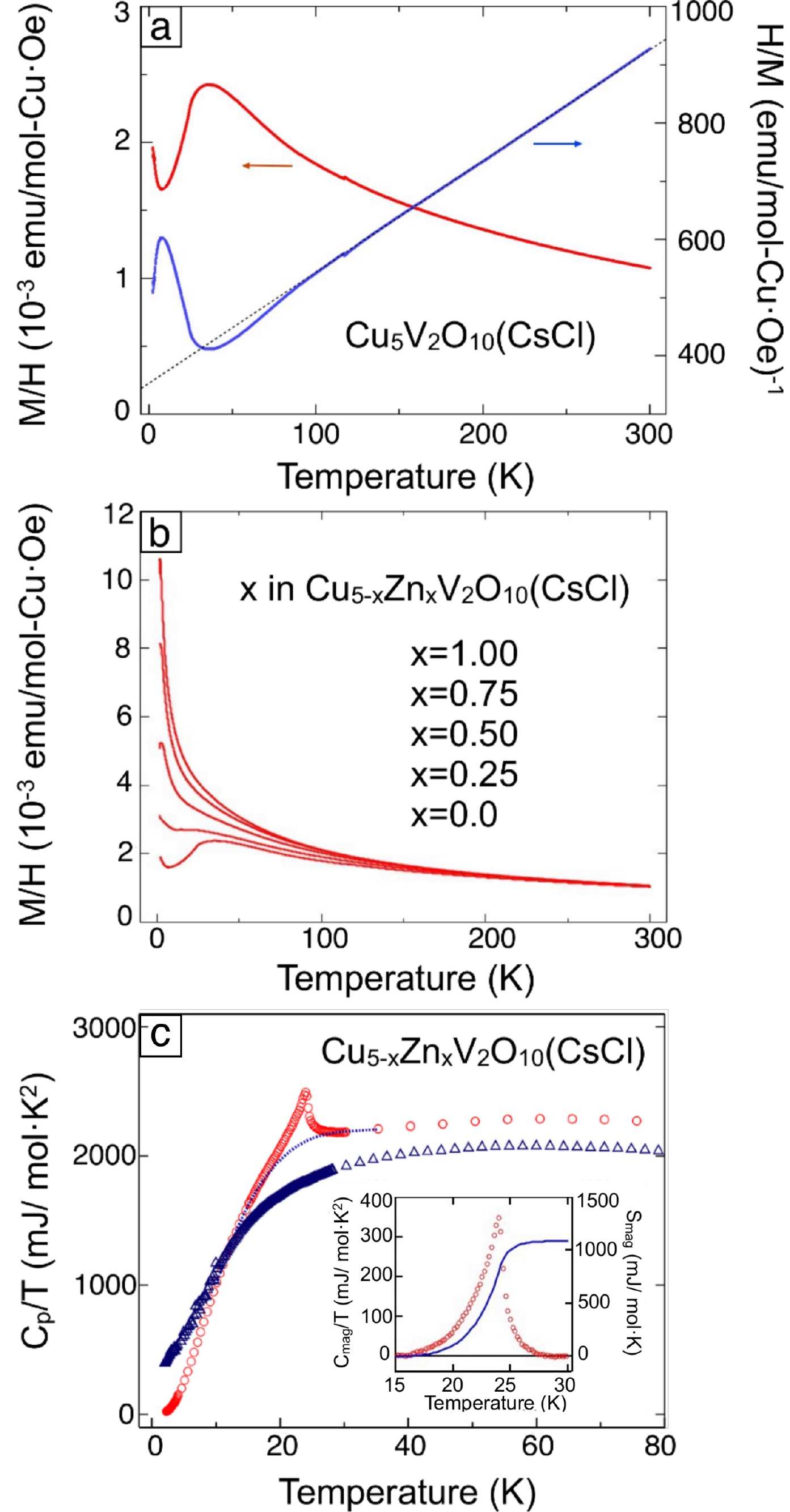}
\caption{(a) DC magnetic susceptibility (left axis) and inverse susceptibility (right axis) of averievite.  The dashed line is a Curie-Weiss fit to the high temperature data. (b) DC magnetic susceptibility of Zn-substituted averievite.  For (a) and (b), data were collected in field-cooled mode (H=2000 Oe). (c) Specific heat of unsubstituted (open circles) and ($x$=1) Zn-substituted averievite (open triangles) plotted as C$_p$/T.  The dotted line is a polynomial background.  Inset: magnetic specific heat (open circles, left axis) and derived magnetic entropy (solid curve, right axis) for averievite.}
\label{fig2}
\end{figure}

We have synthesized the Cs variant of averievite and its Zn-substituted modifications, and via a combination of density functional theory (DFT) calculations
along with magnetic susceptibility and specific heat measurements, unveil the microscopic origin of its electronic and magnetic properties. The susceptibility shows that the parent compound is characterized by an antiferromagnetic (AFM) Curie-Weiss temperature of 185 K (consistent with our theoretical estimate of the superexchange interaction) and orders at 24 K. Experiments show long range order is suppressed by substitution of Cu$^{2+}$ by nonmagnetic Zn, in agreement with our DFT results. Theoretical V$^{5+}$ substitution with Ti$^{4+}$ hole dopes the material and gives rise to an electronic structure near E$_F$ similar to that of a simple kagome tight-binding model, including Dirac points. As such, substituted averievite could allow the realization of novel electronic phases, including not only QSL behavior, but also topological order and unconventional superconductivity.\cite{mazin}

\section{Sample characterization}

As demonstrated by high resolution synchrotron powder diffraction, averievite, Cu$_5$V$_2$O$_{10}$(CsCl), exhibits a complex structural evolution with temperature  (Tables \ref{tablea1}-\ref{tablea3} and Figs.~\ref{figa1}-\ref{figa2} of Appendix A). At 400 K, the data are fully consistent with the space group $P\bar{3}m1$, in agreement with the structure reported by Queen. \cite{queen}  Upon cooling, a phase transition to monoclinic symmetry, P2$_1$/c, is observed at $\sim$ 310 K, reflecting a loss of trigonal symmetry well above T$_N$=24 K. The monoclinic phase is stable down to 127 K, where a second transition occurs to a structure that we have yet to determine.  We note that successful refinement of both trigonal and monoclinic structures requires a disordered Cs site. It is possible that ordering of the Cs ions drives this low-temperature structural phase transition.
Zn-substitution into averievite leads to suppression of both the monoclinic distortion and the structural transition at 127 K (Figs.~\ref{figa3}-\ref{figa4} of Appendix A).

Fig.~\ref{fig2}a shows the temperature dependence of the DC susceptibility $\chi$=M/H of unsubstituted averievite (the data were corrected for core diamagnetism).   Above $\sim$100 K, the susceptibility follows a Curie-Weiss form with an effective moment, p$_{eff}$= 2.05 $\mu_B$/Cu, implying a $g$ factor of 2.36.  The extracted Curie-Weiss temperature $\Theta_{CW}$ of 185 K indicates a dominant antiferromagnetic exchange.  
$\chi$ begins to deviate from this Curie-Weiss behavior below $\sim$100 K, eventually exhibiting a maximum at 35 K, and a low temperature upturn due to uncompensated (orphan) spins (estimated to be at the level of 0.2\%).  A cusp in d$\chi$/dT (not shown) appears at T$_N$=24 K, signaling the onset of long-range antiferromagnetic order, consistent with the behavior of $\chi$ above T$_N$ that is indicative of short range magnetic order.
A frustration index ($\Theta_{CW}$/T$_N$) $\sim$8 implies a moderate degree of magnetic frustration.

Fig.~\ref{fig2}b shows the evolution of the magnetic susceptibility of Zn-substituted averievite Cu$_{5-x}$Zn$_{x}$V$_2$O$_{10}$(CsCl) down to 2 K. The 24 K feature seen in the parent compound is gradually suppressed with increasing Zn content, eventually disappearing at the largest Zn concentration we were able to synthesize without the appearance of second phases, $x$=1. This concentration may represent a solubility limit for Zn substitution under our reaction conditions.
The magnetic behavior is reminiscent of that found in the Zn-paratacamite series Cu$_{4-x}$Zn$_x$(OH)$_6$Cl$_2$ (clinoatacamite $x$=0, herbertsmithite $x$=1). There, Zn substitutes on the interlayer copper sites and magnetic order is also suppressed, with the susceptibility looking remarkably similar between the two $x$=1 compounds.\cite{herb1} Our DFT calculations (see below) as well as crystal chemical considerations argue that Zn$^{2+}$ will substitute preferentially in the honeycomb layers, thus suppressing interlayer magnetic coupling and isolating the kagome planes. Experimental verification of this site selectivity will require neutron diffraction, due to the nearly identical x-ray scattering cross-sections of Cu$^{2+}$ and Zn$^{2+}$ ions.  It may be argued that a magnetic transition is buried under the increasingly large
impurity contribution that is evident as $x$ increases.  However, for $x$=1, there was no evidence for such in $d\chi/dT$.
Moreover, field-cooled and zero field-cooled data for $x$=0 and $x$=1 are equivalent for both 100 Oe and 2000 Oe,
indicating the absence of spin-glass behavior (Fig.~\ref{figa5} of Appendix A).
Lack of evidence for a magnetic transition for $x$=1 is far clearer, though, in the specific heat, as we discuss next.

The specific heat of unsubstituted averievite measured between 1.9 K and 80 K is plotted as C$_p$/T in Fig.~\ref{fig2}c. A prominent $\lambda$-like feature appears in C$_p$/T at T$_N$ = 24 K.  Since a non-magnetic analog is unavailable, we have determined a phenomenological lattice specific heat background as a third-order polynomial fit through the data in the neighborhood of the peak.  The background-subtracted magnetic contribution, C$_{mag}$/T, is shown in the inset of Fig.~\ref{fig2}c along with its integral. The magnetic entropy associated with the transition is calculated as $\Delta$S$_{mag}$ = 1.1 J/mol K or 0.22 J/mol Cu K.  Inevitable errors introduced by the choice of non-magnetic background function are unlikely to have a pronounced impact on this value, which is surprisingly small given the degree of frustration, representing only $\sim$3.8\% of the R ln(2) expected for ordering of S=1/2 spins.  This implies that the bulk of the spectral weight remains fluctuating, with only a small part condensing into the 3D ordered state.  Because of uncertainties in the background subtraction, we did not attempt a scaling analysis of the data near T$_N$.  An unambiguously linear regime in C$_p/$T versus T$^2$ was not found in the temperature range measured.  Nonetheless, extrapolating the lowest temperature data to 0 K suggests that it is gapped. Data below 2 K will be required to corroborate this hypothesis.

Notably, as shown Fig.~\ref{fig2}c, this heat capacity anomaly is not found in $x$=1 Zn-substituted averievite down to the lowest
temperature we have measured (2 K),
consistent with the lack of long range order suggested by the susceptibility.  A plot of C$_p/$T versus T$^2$ (Fig.~\ref{figa6} of Appendix A) is consistent with a large residual
C/T term, likely due to impurities as has been inferred for herbertsmithite.\cite{review_herb}
Besides the gapless behavior evident for $x$=1 at low T, other differences between $x$=0 and $x$=1 are due to the magnetic transition at T$_N$, short range order above T$_N$, and the structural phase transition near 127 K, for $x$=0 (data in Fig.~\ref{fig2}c are shown over a larger temperature range in Fig.~\ref{figa7} of Appendix A).

\section{Computational studies}

\begin{figure}
\includegraphics[width=0.94\columnwidth,draft=false]{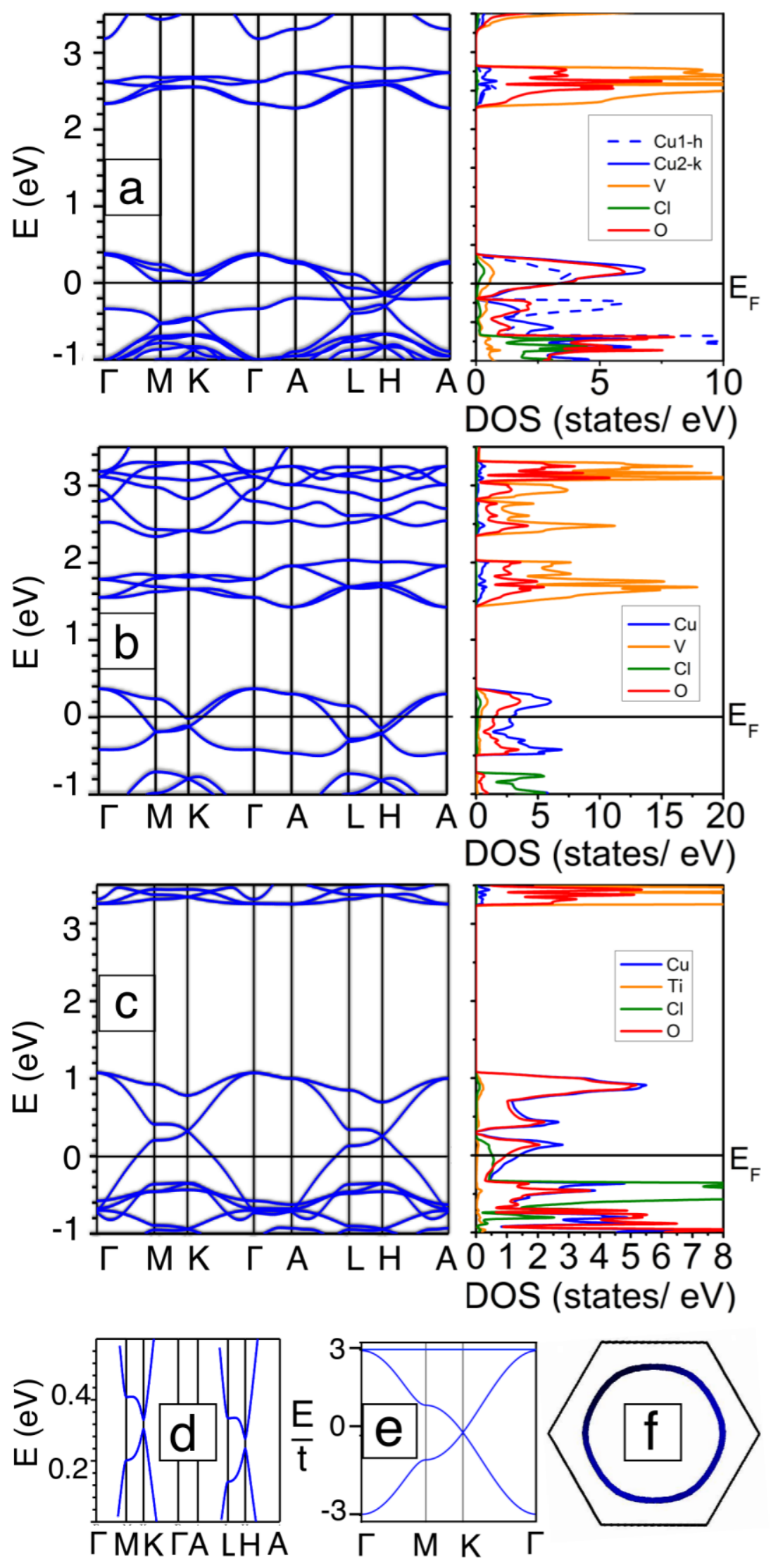}
\caption{GGA band structure and atom-resolved density of states for (a) Cu$_5$V$_2$O$_{10}$(CsCl), (b) Cu$_3$Zn$_2$V$_2$O$_{10}$(CsCl), and (c) Cu$_3$Zn$_2$Ti$_2$O$_{10}$(CsCl). (d) is a blow up around the Dirac points that shows a gap opening due to spin-orbit coupling. (e) Band structure of the nearest-neighbor tight-binding model for the kagome lattice. (f) Fermi surface of Cu$_3$Zn$_2$Ti$_2$O$_{10}$(CsCl).}
\label{fig3}
\end{figure}

We now describe the electronic structure of averievite obtained via DFT calculations performed with the WIEN2k code.\cite{wien2k}  As the monoclinic distortion is very small and suppressed upon Zn-doping, space group $P\bar{3}m1$ was used throughout.
In averievite, within a simple ionic model, one has Cu$^{2+}$ (d$^9$) and V$^{5+}$ (d$^0$).
In the kagome plane, Cu has square planar coordination, so the d$_{x^2-y^2}$ orbitals lie highest in energy. The normal to these CuO$_4$ units sits in the kagome plane and points towards the Cl ions, making averievite different from herbertsmithite. The Cu-O-Cu bond angle in the kagome planes is 115$^{\circ}$. In the honeycomb planes, Cu is surrounded instead by five O with the short Cu-O apical bonds lying along the $c$ axis, so the d$_{z^2}$ orbitals are highest in energy. 
The nonmagnetic GGA atom-resolved density of states (DOS) and band structure are shown in Fig.~\ref{fig3}a and agree with this simple description. A metallic state is obtained with a valence band width of $\sim$ 7 eV (similar to cuprates) and five Cu bands in the vicinity of the Fermi level. The partial DOS were analyzed within the corresponding local coordinate system, showing that the character of these bands is indeed Cu-$d_{x^2-y^2}$ (Cu-kagome) and Cu-$d_{z^2}$ (Cu-honeycomb) as also observed in the charge density (Fig.~\ref{figb1} of Appendix B). 
A large O-$2p$ hybridization can be seen in the DOS (Fig.~\ref{fig3}a).
Given the orientation of the CuO$_4$ units in the kagome plane,
the Cu-d$_{z^2}$ orbitals point towards the Cl ion, leading to hybridization with its $p$ states that appear below -0.7 eV. 
V plays no role near E$_F$ with its unoccupied $d$ states appearing 2 eV above. 

To account for the strong on-site correlations of the Cu-$3d$ electrons, GGA+$U$ \cite{anisimov} calculations were performed with $U$ = 6 eV and $J$ = 1 eV (typical values for cuprates) with a supercell that contains ten inequivalent Cu atoms. The result is an insulating state, and all magnetic configurations explored lead to gaps of $\sim$1.4 eV, similar to other layered cuprates (Fig.~\ref{figb2} of Appendix B). 
The magnetic moment on Cu is $\pm$0.7 $\mu_B$, consistent with S=1/2. The lowest energy solution corresponds to an AFM nearest neighbor (NN) coupling
within the kagome planes, and also between the Cu kagome and Cu honeycomb, resulting in an AFM configuration on a given
Cu tetrahedron. The exchange coupling constants can be obtained by mapping the GGA+$U$ energy differences for different magnetic configurations to a spin 1/2 Heisenberg model. The leading terms are the two NN AFM couplings: $J_1$ connecting Cu-kagome ions, and $J_2$ between Cu-kagome and Cu-honeycomb ions (Fig.~\ref{fig1}). We find that $J_1$ is 203 K and $J_2$ is 35 K. The value of $J_1$ gives rise to a Curie-Weiss temperature similar to experiment.
The ratio of $J_1$ to $J_2$ is 6, similar to the frustration index of 8 from experiment.  Values of $J_n$ for further exchange paths are significantly weaker.

Substitution of the Cu ions in the honeycomb layer with non-magnetic Zn$^{2+}$ should suppress this interlayer coupling. 
Starting from the experimental crystal structure for averievite, all the honeycomb Cu atoms were substituted by Zn in our calculations, giving rise to Cu$_3$Zn$_2$V$_2$O$_{10}$(CsCl). Then, the structure was fully relaxed within GGA until the resulting forces were below 0.01 eV/\AA~(Tables \ref{tableb1}-\ref{tableb2} of Appendix B).  To check for stability, substitution within the kagome planes was also tried.  As the unit cell contains three Cu atoms in the kagome planes, only two of them were substituted by Zn atoms in order to compare the total energies of the two structures. Substitution within the kagome plane turns out to be less stable by 0.8 eV/unit cell, confirming that Zn prefers the trigonal bipyramidal site.

As can be seen from Fig.~\ref{fig3}b, the complex of three Cu-$d_{x^2-y^2}$ 
kagome bands near E$_F$ is well defined and decoupled from other orbitals, due to the removal
of the Cu honeycomb bands. This allows for a tight binding fit of the bands (Fig.~\ref{figb3} and Table \ref{tableb3} of Appendix B).
The resulting exchange interactions can be obtained as before, and a comparable $J_1$ of 170 K is obtained. More importantly, the interlayer coupling is suppressed, given the large separation between the kagome planes (8.5 \AA).  These features combined with the suppression of the magnetic transition upon Zn-doping suggest that Zn-substituted averievite is a promising spin liquid candidate. 

Taking Zn-substituted averievite with isolated kagome planes as a starting point, several doping strategies are apparent.  One could attempt to substitute Zn$^{2+}$ by Ga$^{3+}$ as suggested for herbertsmithite by Mazin {\it et al.}.\cite{mazin}  Or, one could substitute Cs$^{1+}$ by an alkali earth. A promising approach is to substitute V$^{5+}$ with a 4+ ion, Ti$^{4+}$ being the most obvious candidate.
To preserve an undistorted kagome plane, a complete substitution is performed in order to avoid breaking of inversion symmetry.
The resulting formula unit, Cu$_3$Zn$_2$Ti$_2$O$_{10}$(CsCl), corresponds to 2/3 hole doping per Cu. 
Following the same procedure described for Zn-substitution, the structure was fully relaxed. An important outcome is the shift of the kagome oxygen atoms almost completely into the kagome plane (Fig.~\ref{figb1} of Appendix B), which brings the Cu-O-Cu bond angle to almost 120$^{\circ}$ (Tables \ref{tableb1}-\ref{tableb2}  of Appendix B) giving rise to a stronger AFM coupling. 

The resulting electronic structure around the Fermi level (Fig.~\ref{fig3}c) consists of a 3-band complex of Cu-$d_{x^2-y^2}$/O-$p$ character and is qualitatively similar to that of a single orbital model on the kagome lattice. The Cu kagome band complex has a bandwidth that is doubled with respect to the Zn-only substituted case, indicating the extreme sensitivity of the electronic structure to the position of the oxygen ions relative to the kagome plane.
The band dispersion of the three-band complex shows a clear Dirac crossing at $K$ ($K$ and $H$ are almost degenerate, emphasizing the 2D nature of the electronic structure). When spin-orbit coupling is included, a small  gap of 15 meV opens up at the Dirac points (Fig.~\ref{fig3}d). A single orbital kagome tight binding model features symmetry protected Dirac points at $K$ as well as a flat band (Fig.~\ref{fig3}e). Recently, evidence for Dirac fermions in a ferromagnetic kagome metal has been provided.\cite{ye}
Contrary to herbertsmithite, the flat band in averievite lies at the top of the complex rather than the bottom  due to the sign change in
the nearest neighbor hopping parameter (tight binding fits and corresponding hopping parameters can be found in Fig.~\ref{figb3} and Table \ref{tableb3} of Appendix B).\cite{mazin} The Fermi surface is 2D-like
in nature (Fig.~\ref{fig3}f). By doping Ga on the Zn site in a 1:2 ratio, the Fermi level could be raised to the Dirac points  (Fig.~\ref{figb4} of Appendix B).

\section{Summary}

To summarize, averievite opens a new route for the synthesis of
kagome-based structures with the possibility to exhibit unconventional electronic phases with appropriate doping and substitutions.  Our findings exemplify that mineralogical based searches can be a powerful tool for materials design. Given the vast number of known minerals, there is plenty of room for exploration.

\acknowledgments
This work was supported by the  U. S. Department of Energy, Office of Science, Basic Energy Sciences, Materials Sciences and Engineering
Division. Use of the Advanced Photon Source at Argonne National Laboratory was supported by the U. S. Department of Energy, Office of Science, Basic Energy Sciences, under Contract No. DE-AC02-06CH11357. We acknowledge the computing resources provided on Blues, a high-performance computing cluster operated by Argonne's Laboratory Computing Resource Center.

\appendix

\section{Sample synthesis and characterization}

\renewcommand{\thetable}{A\arabic{table}}
\setcounter{table}{0}

\renewcommand{\thefigure}{A\arabic{figure}}
\setcounter{figure}{0}

Polycrystalline samples of averievite (Cu$_5$V$_2$O$_{10}$(CsCl), Cu$_5$-averievite) were synthesized using a conventional solid state reaction. As-delivered CsCl (Johnson-Matthey, 99.9\%), CuO (Johnson-Matthey, 99.999\%), and V$_2$O$_5$ (Alfa-Aesar, 99.995\%) powders were mixed in a 1.048:5:1 ratio. For Zn substituted samples, ZnO (Johnson-Matthey, 99.999\%) was added in stoichiometric proportion.  Excess CsCl was added in all cases to compensate for volatilization during synthesis. Thoroughly ground mixtures were sintered in air at 500$^\circ$C for 12 hours then cooled to 450$^\circ$C at 30$^\circ$C/hour, and then furnace cooled to room temperature. This powder was then pressed into a pellet and sintered again using the same heating cycle as above. 

Laboratory x-ray powder diffraction on the as-prepared samples (PANalytical X'Pert Pro, Cu-K$\alpha$ radiation) showed nearly single phase specimens with a small ($\sim$1\% by Rietveld refinement) contribution from residual CsCl.  Although the CsCl impurity can be washed away with water to yield a 100\% pure product, the magnetic susceptibility of thus purified samples differs from that of the virgin material. Thus, further characterization was restricted to as-prepared specimens. Laboratory x-ray data from both Cu$_5$-averievite and Cu$_4$ZnV$_2$O$_{10}$(CsCl) (Cu$_4$Zn-averievite) samples can be indexed in space group $P\bar{3}m1$, as reported by Queen.\cite{queen}

To further characterize the structure of these samples, high resolution synchrotron powder diffraction data were collected at 30 keV using beamline 11-BM at the Advanced Photon Source (APS), Argonne National Laboratory. An Oxford Cryosystems Cryostream Plus was utilized for sample temperatures over the range 90-450 K. An Oxford Helium Flow cryostat was utilized for measurement of samples at 15 K. Samples were indexed, solved, and refined by the Rietveld method utilizing TOPAS v5 (Bruker) and ISODistort.\cite{campbell}

\begin{figure}
\center
\includegraphics[width=0.85\columnwidth,draft=false]{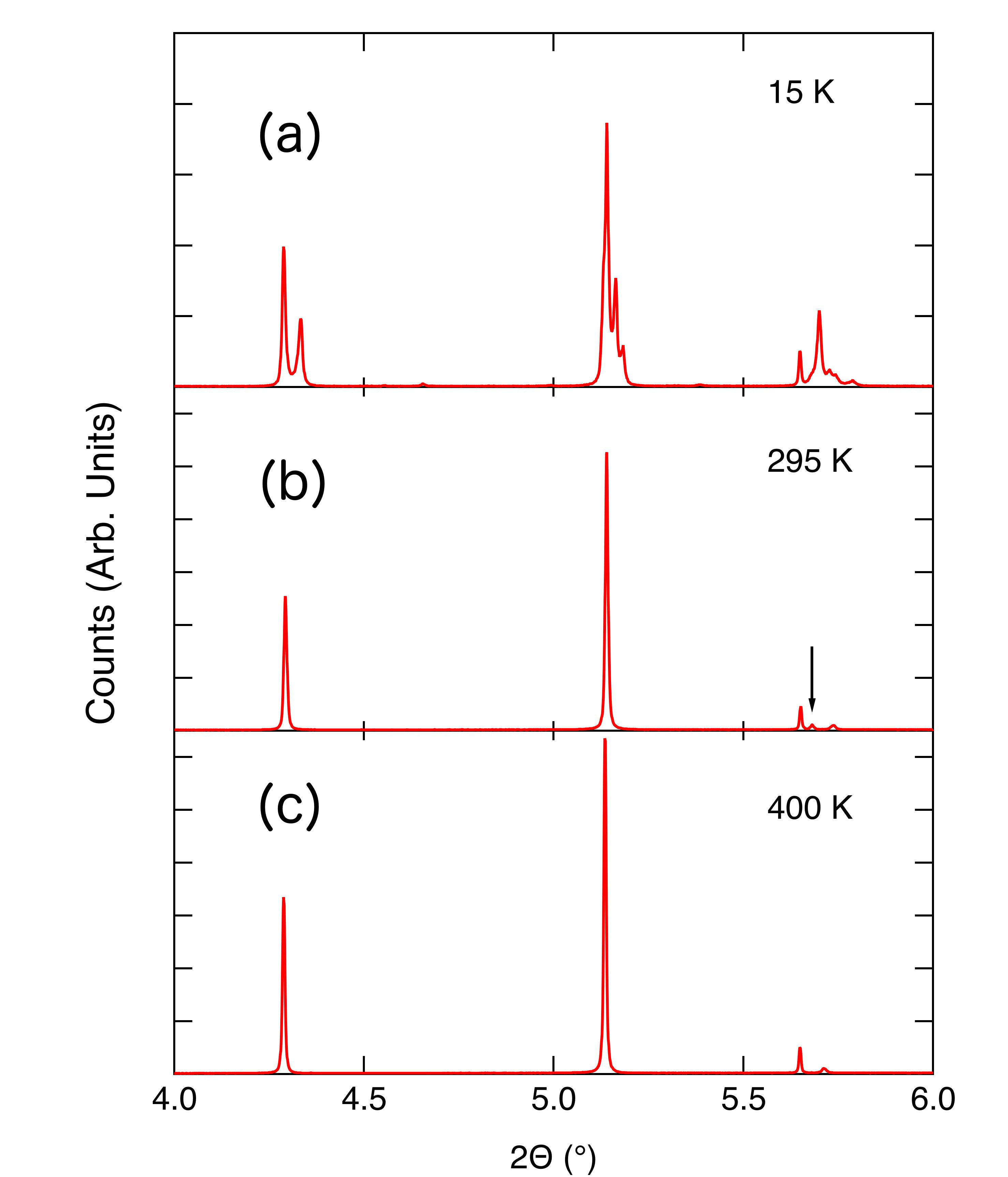}
\caption{Synchrotron X-ray powder diffraction data for Cu$_5$-averievite measured at 15 K, 295 K, and 400 K.}
\label{figa1}
\end{figure}

\begin{figure}
\center
\includegraphics[width=0.8\columnwidth,draft=false]{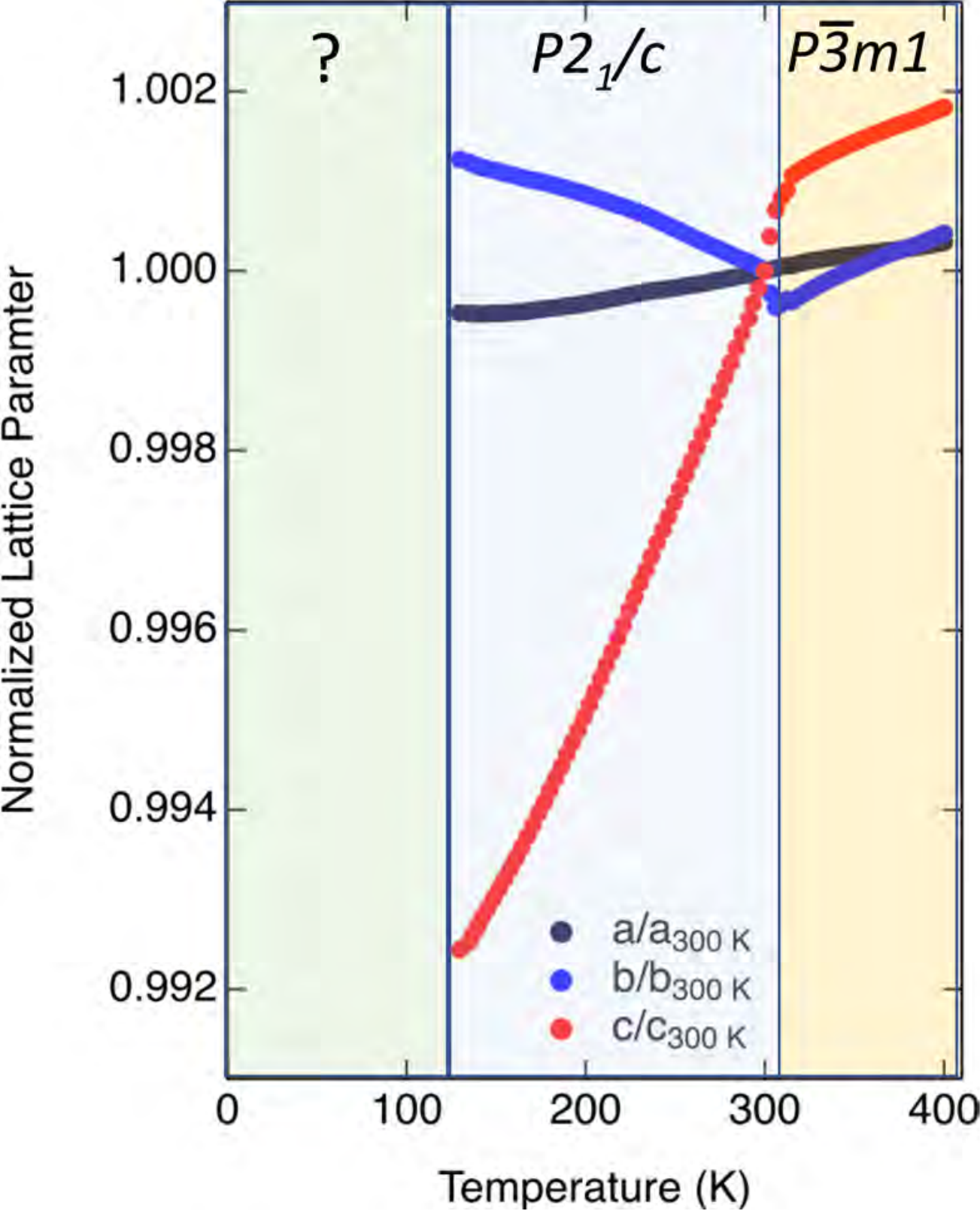}
\caption{Temperature dependence of lattice parameters for Cu$_5$-averievite normalized to 300 K as determined by high resolution synchrotron x-ray powder diffraction.  In the $P\bar{3}m1$ regime above 310 K, the lattice is described in terms of an equivalent monoclinic cell with $a'$= $c$, $b'$= $a$, $c'$= a$\sqrt3$ to facilitate comparison with the low temperature phase. The structure below 127 K was not indexed.  See text for details.}
\label{figa2}
\end{figure}

The data measured at 400 K for Cu$_5$-averievite, shown in Fig.~\ref{figa1}(c), can be successfully indexed and refined in the trigonal $P\bar{3}m1$ structure. However, as shown in Fig.~\ref{figa1}(b), at 295 K a small peak appears near 2$\theta$ = 5.7$^{\circ}$, which is absent at 400 K, reflecting a lower symmetry indexed to monoclinic $P2_{1}/c$.  Refined lattice parameters at these two temperatures are shown in Table \ref{tablea1}.  Crystallographic data for the monoclinic phase at 295 K and the trigonal phase at 400 K are presented in Table \ref{tablea2} and Table \ref{tablea3}, respectively.  Notably, the Cs positions both in the trigonal and monoclinic phases are disordered across two symmetry-equivalent sites with statistical occupancy.  Fig.~\ref{figa2} shows the temperature dependence of the lattice constants, revealing that the trigonal-monoclinic transition occurs at ~310 K.  Upon further cooling, a second structural phase transition occurs at 127 K (also seen as a small anomaly in the heat capacity).  As shown by the data in Fig.~\ref{figa1}(a), a complex superstructure is likely, potentially incommensurate with the underlying monoclinic subcell, as no simple multiples of this subcell successfully index the pattern.  It is plausible that this transition arises from an ordering of the Cs ions along the $c$-axis, but solution of the structure lies outside the scope of this work.

\begin{table}
\caption{Crystal structure for Cu$_5$-averievite Cu$_5$V$_2$O$_{10}$(CsCl) at 400 K and 295 K}
\begin{ruledtabular}
\begin{tabular}{ccc}
\multicolumn{1}{c}{Temperature (K)} &
\multicolumn{1}{c}{400} &
\multicolumn{1}{c}{295}   \\
\hline
System &  Trigonal & Monoclinic     \\ 
Space group & $P\bar3m1$  & $P2_1/c$    \\ 
$a$ (\AA) & 6.369326(10)   & 8.373335(13)   \\     
$b$ (\AA) & 6.369326(10) & 6.366556(14)   \\
$c$ (\AA) & 8.375817(12) & 11.01280(3) \\
$\alpha$ & 90 & 90  \\
$\beta$ & 90 & 90.0221(4)  \\
$\gamma$ & 120 & 90  \\
Volume (\AA$^3$) & 294.269(1) & 587.084(2)  \\
R$_{WP}$ & 9.39 & 8.68 \\
R$_{exp}$ & 5.96 & 3.96 

\label{tablea1}
\end{tabular}
\end{ruledtabular}
\end{table}

\begin{table*}
\caption{Atomic coordinates for Cu$_5$-averievite at 295 K (space group $P2_{1}/c$)}
\begin{ruledtabular}
\begin{tabular}{lcccccc}
\multicolumn{1}{l}{lattice constants (\AA)} &
\multicolumn{1}{c}{$x$} &
\multicolumn{1}{c}{$y$} & 
\multicolumn{1}{c}{$z$} & 
\multicolumn{1}{c}{$B_{iso}$} &
\multicolumn{1}{c}{Occupancy} & \\
\hline
   Cs &  0.44984(8) & 0.0005(9)  & 0.9998(3) & 3.8110(18) & 0.5\\ 
  Cu$_h$ & 0.72932(8) & 0.5077(3) & 0.16639(15) & 1.213(9) & 1.0 \\
   Cu$_{k1}$ &  0.9999(3) & 0.77247(12)  & 0.23899(7) & 1.213(9) & 1.0\\
      Cu$_{k2}$ & 0 & 0 & 0.5 & 1.213(9) &  1.0 \\
   V &  0.3050(4) & 0.5096(4) & 0.1669(2) & 0.579(16)  & 1.0 \\
    Cl & 0  & 0  & 0 & 3.66(4) & 1.0 \\
  O$_{Cs}$ &  0.5007(5) & 0.5167(5)  & 0.1685(5)& 0.579(16) & 1.0 \\
 O$_k$ &  0.9507(3) & 0.5096(7)  & 0.1637(5) & 0.579(16) & 1.0 \\
  O$_{h1}$ &  0.7626(7) & 0.6929(7) & 0.7441(4) & 0.579(16) & 1.0 \\
    O$_{h2}$ &  0.7627(7) & 0.2254(7) & 0.2498(4)& 0.579(16) & 1.0 \\
      O$_{h3}$ &  0.7545(7)& 0.0032(13)& 0.4751(5)& 0.579(16) & 1.0\\
\label{tablea2}
\end{tabular}
\end{ruledtabular}
\end{table*}

\begin{table*}
\caption{Atomic coordinates for Cu$_5$-averievite at 400 K (space group $P\bar{3}m1$)}
\begin{ruledtabular}
\begin{tabular}{lcccccc}
\multicolumn{1}{l}{lattice constants (\AA)} &
\multicolumn{1}{c}{$x$} &
\multicolumn{1}{c}{$y$} & 
\multicolumn{1}{c}{$z$} & 
\multicolumn{1}{c}{$B_{iso}$} & 
\multicolumn{1}{c}{Occupancy} & \\
\hline
   Cs &  0 & 0 & 0.54965 & 4.24(5) & 0.5 \\ 
  Cu$_h$ & 1/3  & 2/3  & 0.72936 & 1.754(10) & 1.0\\
   Cu$_k$ &  1/2 & 1/2  & 0 & 1.754(10) & 1.0 \\
   V &  1/3 & 2/3  & 0.30485(12) & 0.816(16) & 1.0\\
    Cl & 0  & 0  & 0 & 4.53(2) & 1.0\\
  O$_{Cs}$ &  1/3 & 2/3  & 0.5011(6) & 0.816(16) & 1.0\\
 O$_k$ &  1/3 & 2/3  & 0.9502(4) & 0.816(16) & 1.0\\
  O$_h$ &  0.47713(17) & 0.52287(17) & 0.2410(2) & 0.816(16) & 1.0\\
\label{tablea3}
\end{tabular}
\end{ruledtabular}
\end{table*}

\begin{figure}
\center
\includegraphics[width=\columnwidth,draft=false]{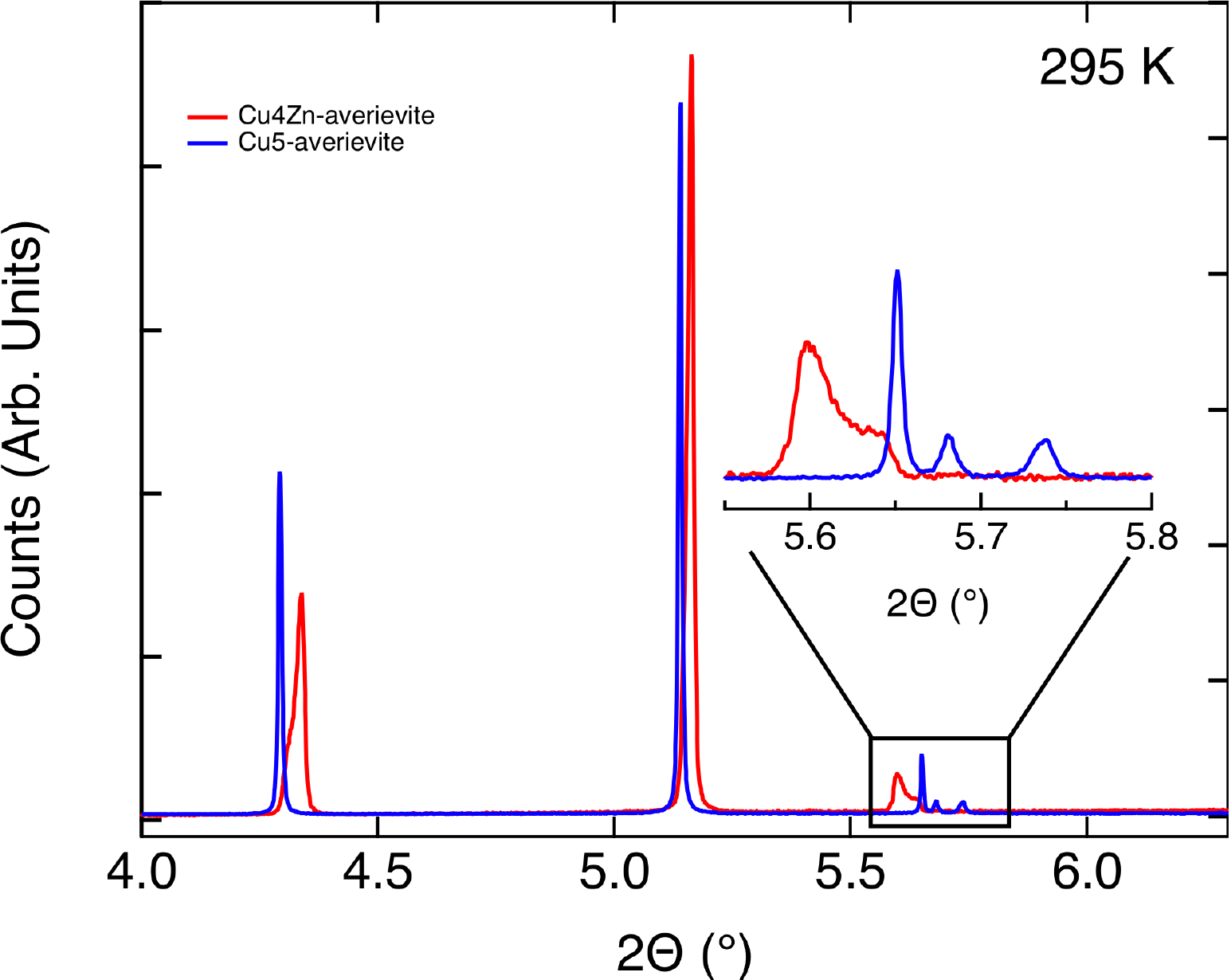}
\caption{Synchrotron x-ray powder diffraction data for Cu$_5$- and Cu$_4$Zn-averievite measured at 295 K.}
\label{figa3}
\end{figure}

\begin{figure}
\center
\includegraphics[width=\columnwidth,draft=false]{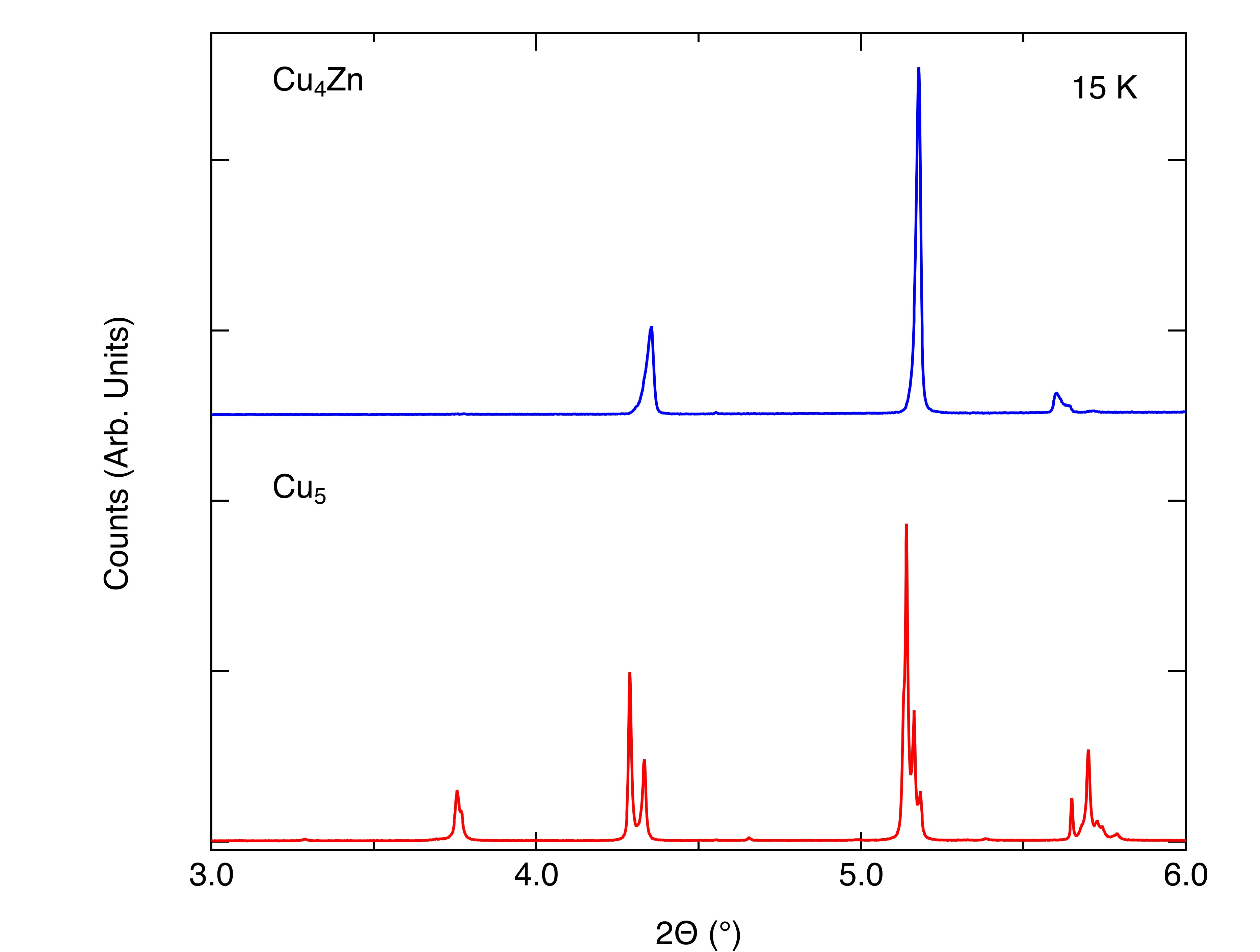}
\caption{Synchrotron x-ray powder diffraction data for Cu$_4$Zn- and Cu$_5$-averievite at 15 K.}
\label{figa4}
\end{figure}

Synchrotron data measured on Cu$_4$Zn-averievite show weak shoulders on every peak, suggesting that our samples are biphasic (Fig.~\ref{figa3}).  It is unclear if this reflects a true thermodynamic phase separation or incomplete homogenization of the sample during synthesis at modest temperatures.  Importantly, based on the diffraction data shown in Fig.~\ref{figa3}, neither component of the biphasic mixture corresponds to Cu$_5$-averievite; that is, each has a significant degree of Zn substitution.  Notably, the data show no peak at 2$\theta$ = 5.7$^\circ$ at any temperature, nor is the low-temperature transition found in Cu$_5$-averievite observed (Fig.~\ref{figa4}).  These results imply that the nominal Cu$_4$Zn phase is either trigonal throughout the measured temperature range or has an extremely weak monoclinic distortion.

\begin{figure}
\center
\includegraphics[width=\columnwidth,draft=false]{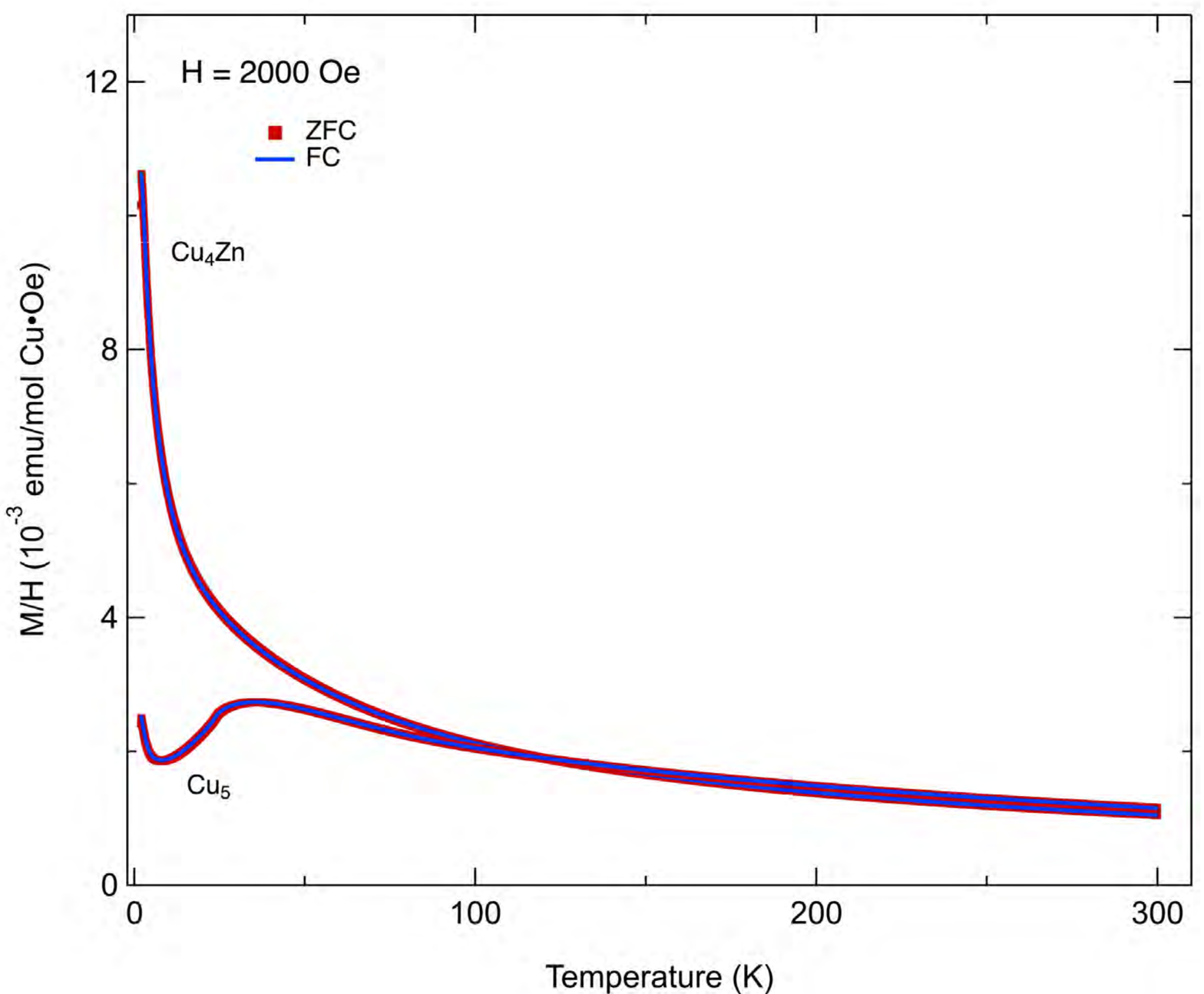}
\caption{Zero field-cooled and field-cooled magnetic susceptibility at 2000 Oe for Cu$_5$- and Cu$_4$Zn-averievite.}
\label{figa5}
\end{figure}

To explore the possibility of spin glass behavior, we measured the magnetic susceptibility as a function of temperature using both zero field-cooled (ZFC) and field-cooled (FC) protocols on samples of Cu$_5$-averievite and Cu$_4$Zn-averievite.   The results for a measuring field of H=2000 Oe are shown in Fig.~\ref{figa5}.  In both the unsubstituted and Zn-substituted averievite, no difference is seen between the FC and ZFC curves. The same reversible behavior was found at H=100 Oe.  The observed reversibility argues against a glassy magnetic state.

\begin{figure}
\center
\includegraphics[width=\columnwidth,draft=false]{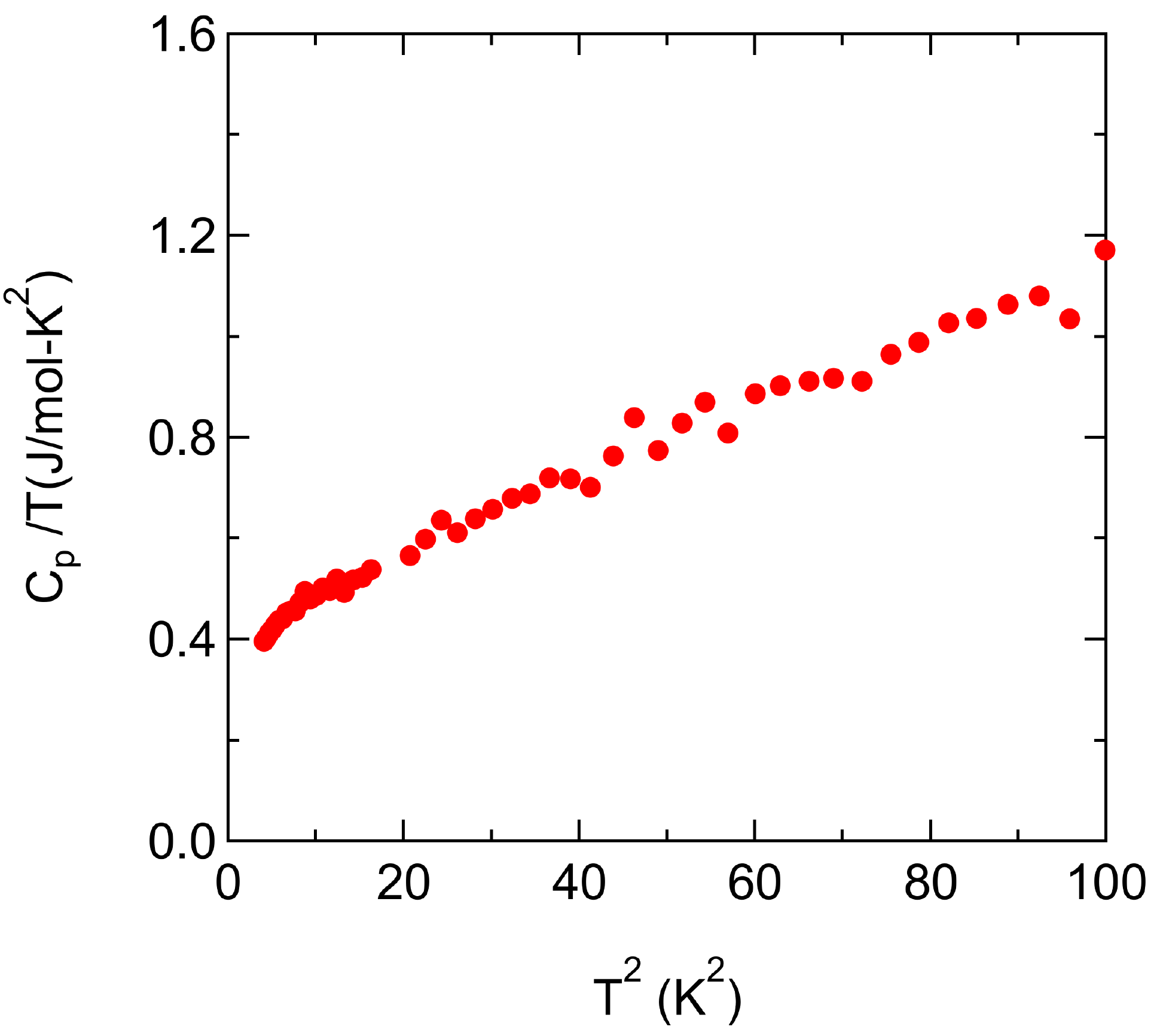}
\caption{Heat capacity of Cu$_4$Zn-averievite below 10 K.}
\label{figa6}
\end{figure}

Fig.~\ref{figa6} shows the measured heat capacity of a sample of Cu$_4$Zn-averievite below 10 K plotted as C$_p$/T vs. T$^2$.  We note that extrapolation to T=0 yields a finite intercept of 400 mJ/mol-K$^2$.  Such a T-linear component implies gapless magnetic excitations, with impurity spins being a likely origin.  Indeed, the value reported by Han {\it et al.}~\cite{han} in herbertsmithite and attributed to residual spin-1/2 impurities is similar to that found here.

\begin{figure}
\center
\includegraphics[width=\columnwidth,draft=false]{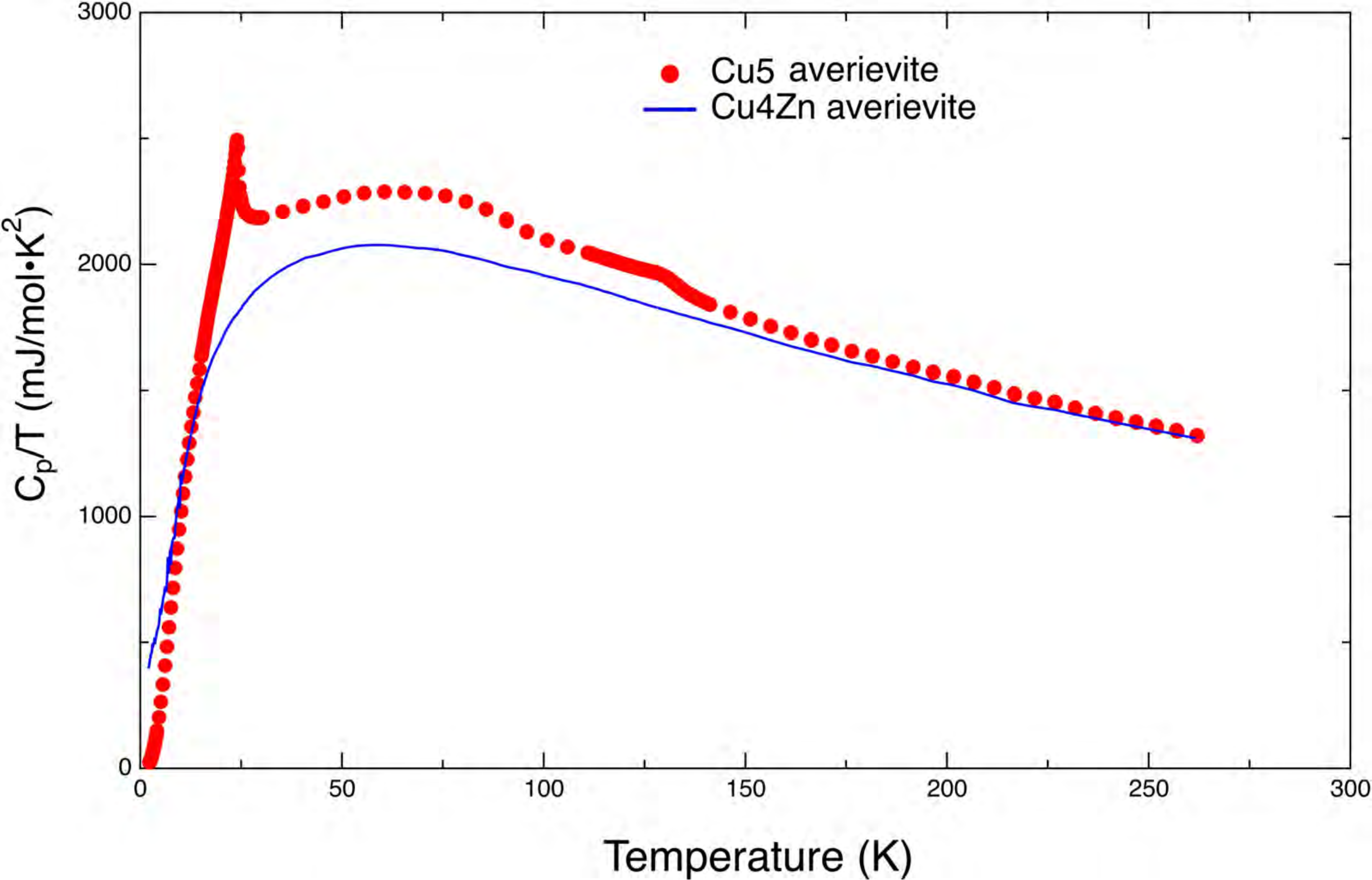}
\caption{Heat capacity of Cu$_5$-averievite and Cu$_4$Zn-averievite between 2 and 262 K.}
\label{figa7}
\end{figure}

Fig.~\ref{figa7} shows the heat capacity of samples of Cu$_5$-averievite and Cu$_4$Zn-averievite plotted measured between 2 and 262 K. Above $\sim$150 K, these curves converge as expected.
However, Cu$_5$-averievite has an enhanced heat capacity relative to the Zn-substituted analog.  We attribute this enhancement to short-range correlations developing among the Cu spins as precursory to the long-range ordered state.  The absence of this behavior in Cu$_4$Zn-averievite is consistent with it being a spin liquid.
In addition, there is a weak anomaly in Cu$_5$-averievite near 127 K due to the structural phase transition.

\renewcommand{\thefigure}{B\arabic{figure}}
\setcounter{figure}{0}

\renewcommand{\thetable}{B\arabic{table}}
\setcounter{table}{0}

\section{Computational details}

We chose R$_{mt}$K$_{max}$=7.0, and the muffin tin radii (a.u.) were 2.5 for Cs, 1.89
for Cu, 1.85 for Zn, 1.61 for Ti, 1.63 for V,  2.5 for Cl, and 1.5 for O. 
Well converged meshes of 184 $k$-points for the conventional cell and 105 $k$-points for the supercell in the
irreducible wedge of the zone were used.

The structural details for calculated structures for Cu$_3$Zn$_2$V$_2$O$_{10}$(CsCl) (Cu$_3$Zn$_2$-averievite) and  Cu$_3$Zn$_2$Ti$_2$O$_{10}$(CsCl) (Cu$_3$Zn$_2$Ti$_2$-averievite) are described in Tables \ref{tableb1} and \ref{tableb2}.

\begin{figure}
\center
\includegraphics[width=0.8\columnwidth,draft=false]{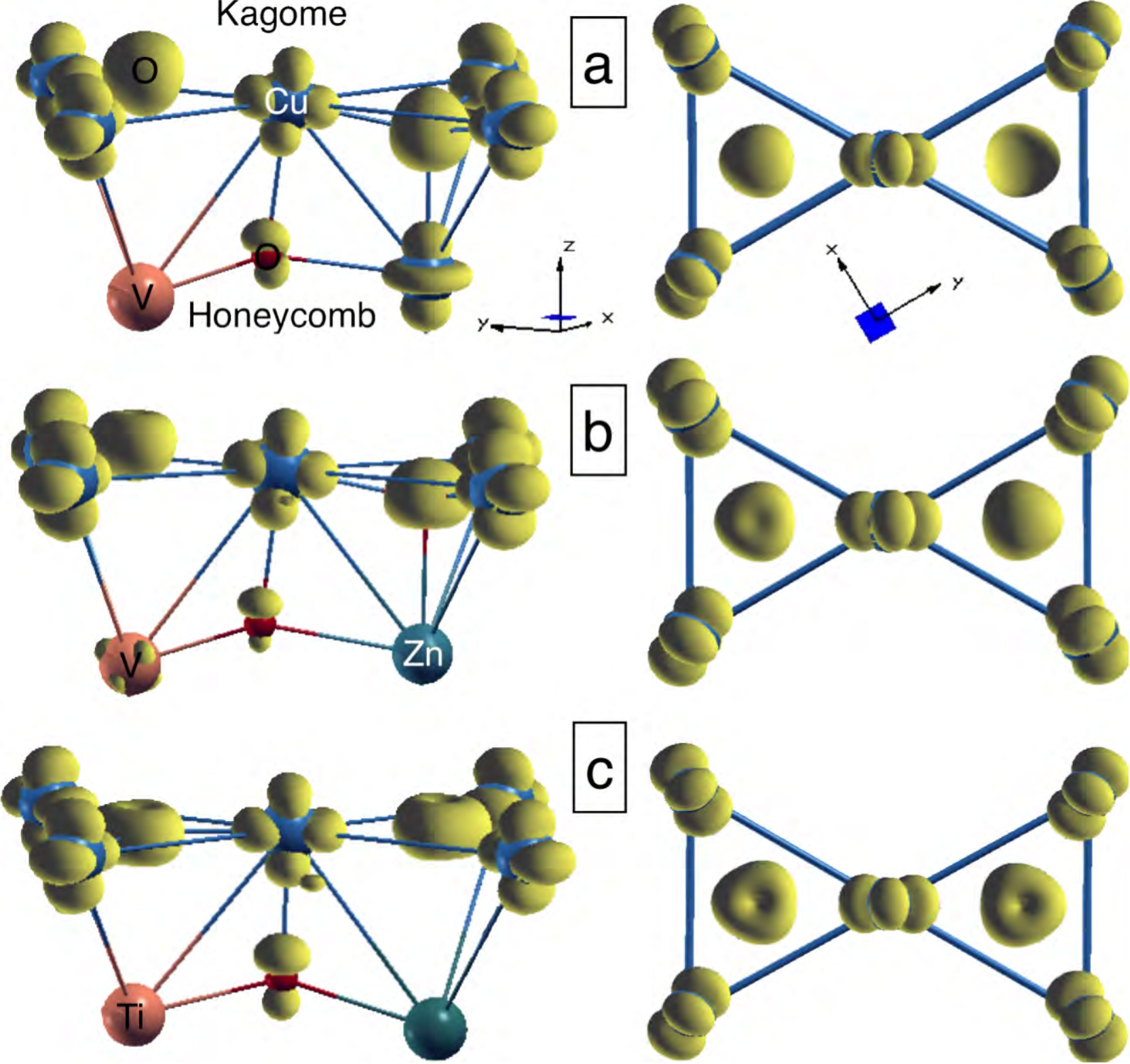}
\caption{Charge density within GGA in an energy window that comprises the three kagome Cu d bands around E$_F$ for (a) Cu$_5$V$_2$O$_{10}$(CsCl), (b) Cu$_3$Zn$_2$V$_2$O$_{10}$(CsCl), and (c) Cu$_3$Zn$_2$Ti$_2$O$_{10}$(CsCl).}\label{figb1}
\end{figure}

\begin{table*}
\caption{Calculated crystal structure for Cu$_3$Zn$_2$-averievite, Cu$_3$Zn$_2$Ti$_2$-averievite with space group P$\bar{3}$m1}
\begin{ruledtabular}
\begin{tabular}{lccccc}
\multicolumn{1}{l}{lattice constants (\AA)} &
\multicolumn{1}{c}{$a$} &
\multicolumn{1}{c}{$b$} & 
\multicolumn{1}{c}{$c$} &  \\
\hline
Cu$_3$Zn$_2$-averievite &  6.5116 & 6.5116 &   8.5666  &     \\
Cu$_3$Zn$_2$Ti$_2$-averievite & 6.5754 & 6.5754 &  8.6505&      \\
\hline
\hline
      atomic positions & x$_{Cu_3Zn_2}$/ x$_{Cu_3Zn_2Ti_2}$ &  y$_{Cu_3Zn_2}$/ y$_{Cu_3Zn_2Ti_2}$ & z$_{Cu_3Zn_2}$/ z$_{Cu_3Zn_2Ti_2}$ \\    
\hline
   Cs &  0 & 0  & 1/2 \\ 
  Cu$_h$ & 1/3  & 2/3  &  0.7236/ 0.7060\\
   Cu$_k$ &  1/2 & 1/2  & 0 \\
   V/Ti &  1/3 & 2/3  &  0.2998/ 0.2746\\
    Cl & 0  & 0  & 0\\
  O$_{Cs}$ &  1/3 & 2/3  & 0.5028/ 0.5221 \\
 O$_k$ &  1/3 & 2/3  & 0.9582/ 0.9957\\
  O$_h$ &   0.4809/ 0.4961 & 0.5191/ 0.5039 & 0.2355/ 0.2200
\label{tableb1}
\end{tabular}
\end{ruledtabular}
\end{table*}

\begin{table*}
\caption{Bond angles between transition metal and oxygen ions, and nearest neighbor (NN) distances for Cu$_5$-averievite (experimental), Cu$_3$Zn$_2$-averievite (calculated), and Cu$_3$Zn$_2$Ti$_2$-averievite (calculated). Between parenthesis the number of NN at each distance is specified.}
\begin{ruledtabular}
\begin{tabular}{lccccc}
\multicolumn{1}{l}{angles ($^\circ$)} &
\multicolumn{1}{c}{Cu$_k$-O-Cu$_k$} &
\multicolumn{1}{c}{O$_h$-Cu$_k$-O$_k$} & 
\multicolumn{1}{c}{  O-Zn/Cu$_h$-O} & 
\multicolumn{1}{c}{O-V/Ti-O} & \\
\hline
Cu$_5$-averievite &  115 & 84  &  118 & 110   \\
Cu$_3$Zn$_2$-averievite &  117 & 85  &  117 & 110   \\
Cu$_3$Zn$_2$Ti$_2$-averievite &  120 & 88  &  110 & 114   \\
\hline
\hline
NN distance (\AA) &   & Cu$_k$-O & Zn/Cu$_h$-O & V/Ti-O   \\
\hline
Cu$_5$-averievite &  in-plane &  1.88(2) & 2.08(3) &  1.7(3) &  \\
Cu$_5$-averievite &  out-of-plane &  2.04(2) & 1.89(1), 1.87(1) &  1.64(1), 2.95(1) &  \\
Cu$_3$Zn$_2$-averievite &  in-plane &  1.91(2) & 2.12(3) &  1.75(3) &  \\
Cu$_3$Zn$_2$-averievite &  out-of-plane &  2.04(2) & 1.94(1), 2.00(1) &  1.69(1), 2.91(1) &  \\
Cu$_3$Zn$_2$Ti$_2$-averievite &  in-plane &  1.89(2) & 2.04(3) &  1.91 &  \\
Cu$_3$Zn$_2$Ti$_2$-averievite &  out-of-plane &  2.04(2) & 1.97(1), 2.58(1) &  1.69(1), 2.33(1) &  
\label{tableb2}
\end{tabular}
\end{ruledtabular}
\end{table*}

\begin{table}
\caption{Relevant hopping parameters in meV from tight binding fits for Cu$_3$Zn$_2$-averievite (AV) and Cu$_3$Zn$_2$Ti$_2$-averievite (AV) compared to those of herbertsmithite (HERB), kapellasite (KAP) and haydeeite (HAY).\cite{janson} Note that the definition of $t_p$ and its associated $d$ varies depending on the crystal structure.}
\begin{ruledtabular}
\begin{tabular}{lcccccc}
\multicolumn{1}{l}{$t$} &
\multicolumn{1}{c}{Cu$_3$Zn$_2$-AV} &
\multicolumn{1}{c}{Cu$_3$Zn$_2$Ti$_2$-AV} & 
\multicolumn{1}{c}{ HERB} & 
\multicolumn{1}{c}{KAP} & 
\multicolumn{1}{c}{HAY} & \\
\hline

\hline
$t_1$ &  -113 &  -221 & 163 &  87 & 73  \\
$t_2$ &  -27 &  -42  & -  &  -10  &  -9 \\
$t_{3d2}$ &  41 &  18 & 23  &  20 & 22  \\
$t_{3d}$ &  -14 &  -21 &-  &  49 &  42 \\
$t_p$ &  22 &  16 &  37 & -  & - 

\label{tableb3}
\end{tabular}
\end{ruledtabular}
\end{table}

\begin{figure*}
\center
\includegraphics[width=0.9\columnwidth,draft=false]{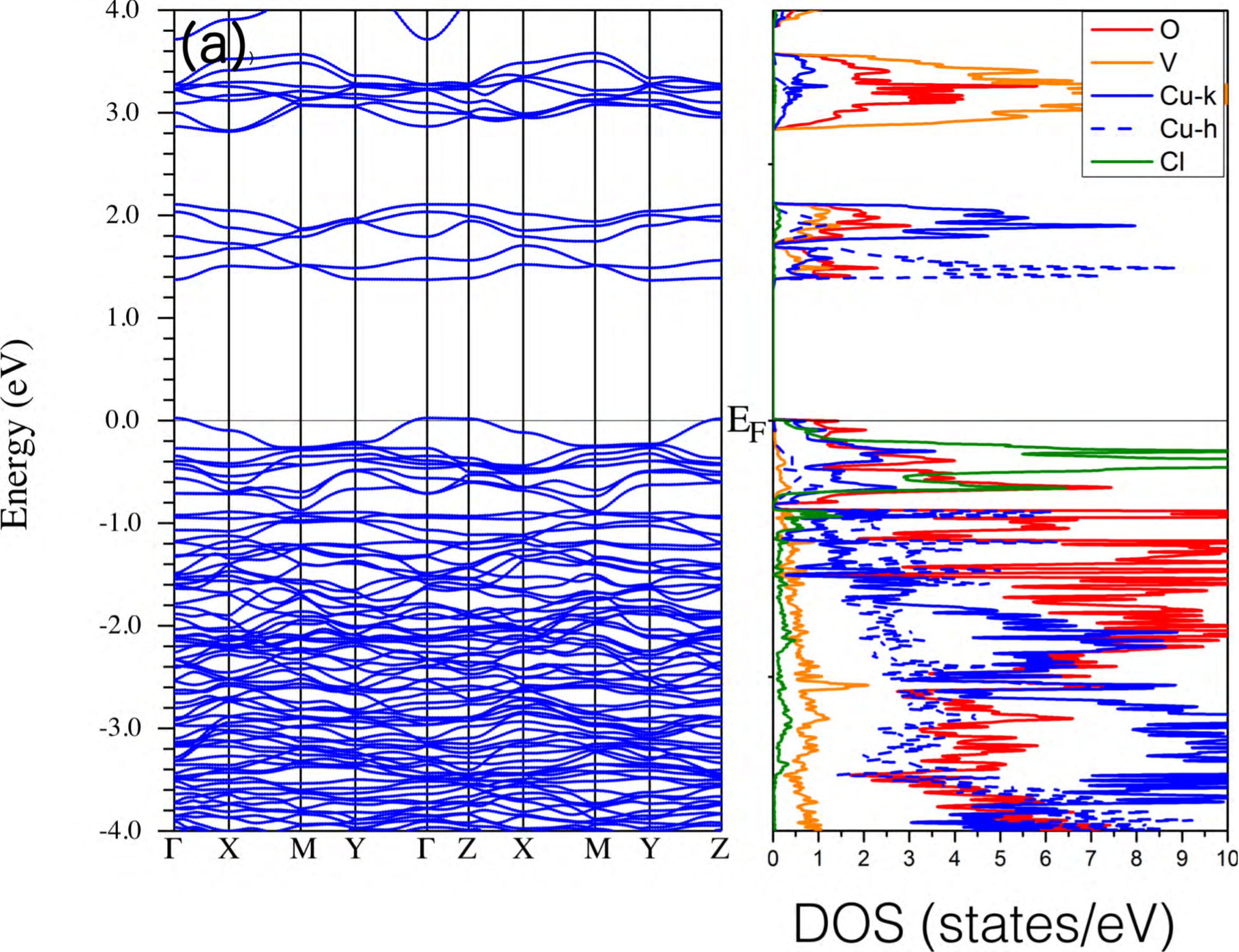}
\includegraphics[width=0.9\columnwidth,draft=false]{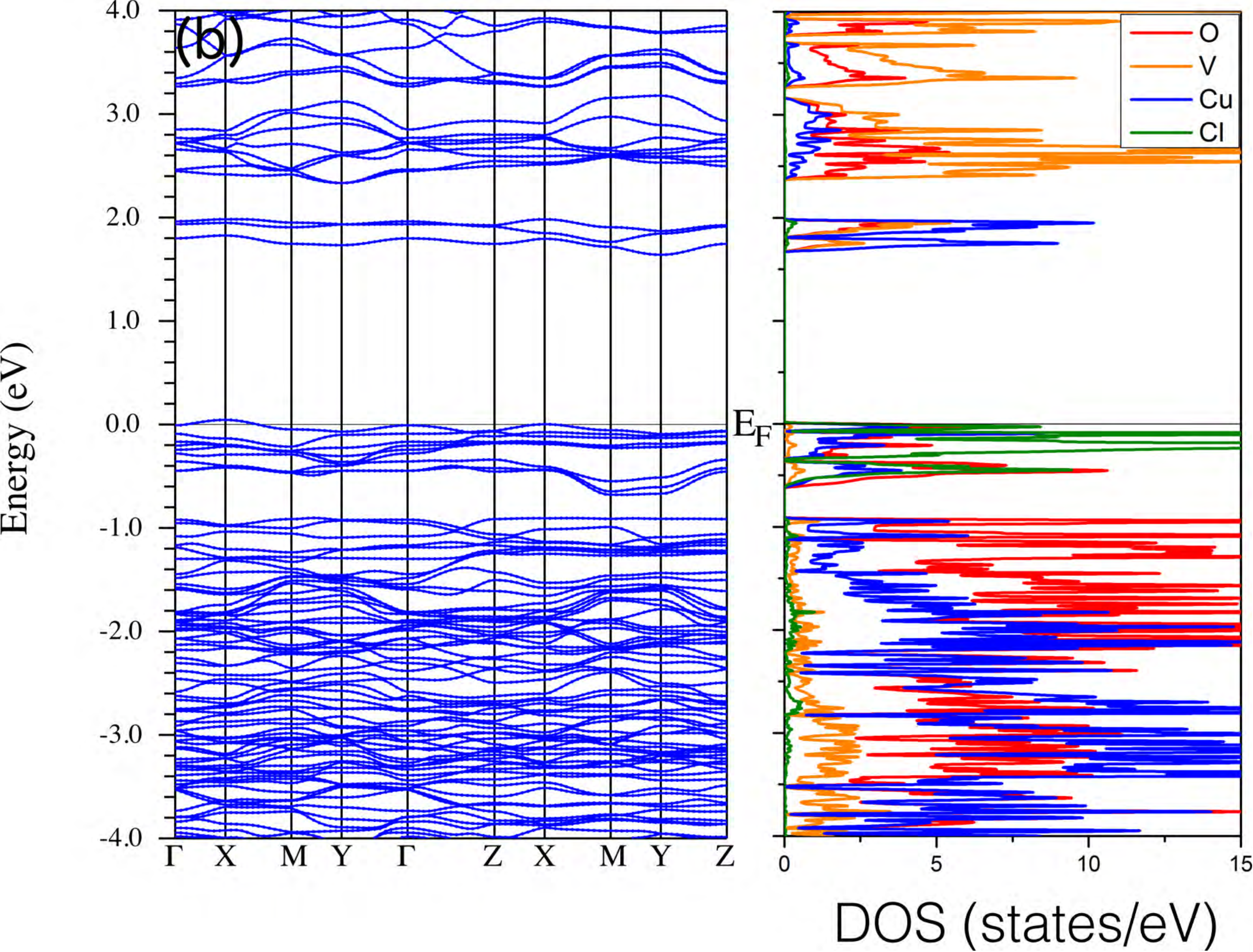}
\caption{GGA+$U$ AFM band structure and atom resolved density of states for (a) Cu$_5$-averievite (b) and Cu$_3$Zn$_2$-averievite.}\label{figb2}
\end{figure*}

DFT+$U$ calculations were performed in a supercell with ten inequivalent Cu atoms to allow
for the determination of various exchange parameters, obtained from energy differences of various magnetic configurations
mapped onto the analogous Heisenberg model.  The GGA+$U$ AFM band structure and DOS for Cu$_5$-averievite (Fig.~\ref{figb2}(a)) show the material becomes insulating once magnetism and on-site correlations are considered.  The gap is 1.4 eV, with the Cu-honeycomb-d$_{z
^2}$ states and Cu-kagome-d$_{x^2-y^2}$ states partly unoccupied.  At higher energies ($\sim$3 eV), the unoccupied V-$d$ states show up.
The energy spectrum right below the Fermi level is dominated by Cl-$p$ states hybridized with Cu-$d$. Given the particular environment of the Cu atoms in the kagome plane, its Cu-d$_{z^2}$ orbitals lie within it and point directly towards
the Cl ion, leading to this strong degree of hybridization. Most of the occupied Cu-$d$ and O-$p$ weight appears below -0.8 eV.  The magnetic moments inside the Cu muffin tin spheres are $\sim$0.7$\mu_B$, consistent with S=1/2. No sizable moments develop on the oxygen atoms.

\begin{figure*}
\center
\includegraphics[width=1.8\columnwidth,draft=false]{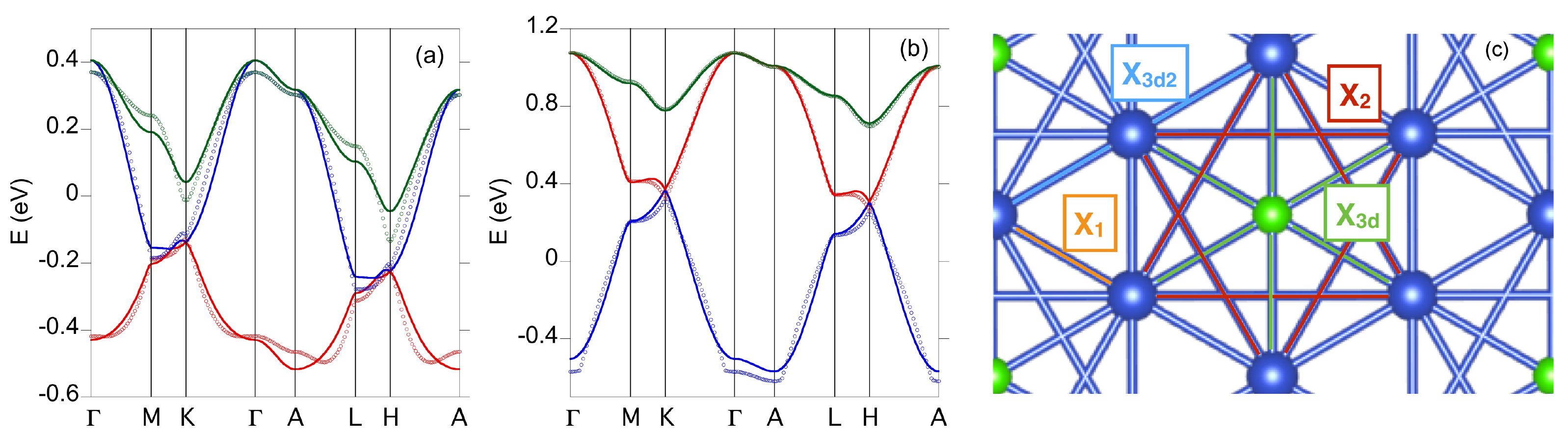}
\caption{ GGA band structure and tight-binding fits corresponding to (a) Cu$_3$Zn$_2$-averievite and (b) Cu$_3$Zn$_2$Ti$_2$-averievite. (c) Exchange paths within the kagome plane associated with the hopping parameters shown in Table \ref{tableb3}. Cu atoms are shown in blue, Cl atoms in green.}\label{figb3}
\end{figure*}

\begin{figure}
\center
\includegraphics[width=0.8\columnwidth,draft=false]{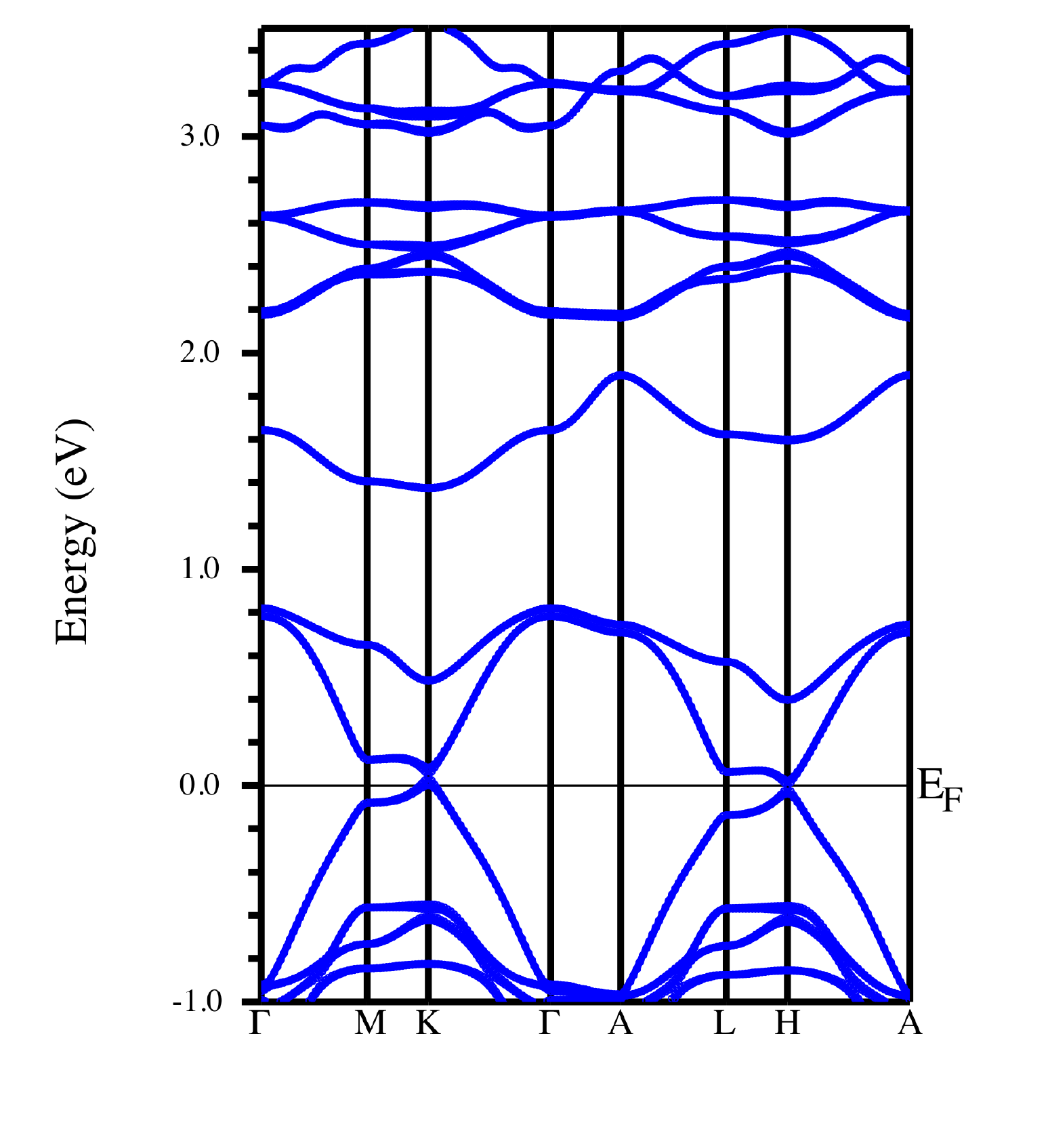}
\caption{Band structure of Cu$_3$ZnGaTi$_2$O$_{10}$(CsCl).}\label{figb4}
\end{figure}

Fig.~\ref{figb2}(b) shows the GGA+$U$ AFM band structure and DOS for Cu$_3$Zn$_2$-averievite. The picture is similar to that shown in Fig.~\ref{figb2}(a) for averievite, other than the obvious removal of the Cu-$d$ honeycomb states, leaving the three unoccupied Cu-d$_{x^2-y^2}$ kagome (minority spin) bands isolated.
The band gap with respect to averievite increases from 1.4 eV to 1.7 eV, and the unoccupied V-$d$ states appear at a slightly lower energy. The magnetic moments inside the Cu muffin tin spheres are the same at $\sim$0.7$\mu_B$. 

%\subsection{Tight binding fits}

To obtain useful parameters for future work on these materials, a simple three-band tight binding model was fit to the GGA bands shown in Figs.~3(b) and 3(c) of the main text (corresponding to Cu$_3$Zn$_2$-averievite and Cu$_3$Zn$_2$Ti$_2$-averievite, respectively). Table \ref{tableb3} shows the corresponding hopping parameters ($t$) compared to those for other kagome lattice compounds.\cite{janson}  For averievite, the relevant terms are the nearest-neighbor (NN) coupling ($t_1$, $d$=$a/2$), the next NN coupling ($t_2$, $d=a/(2\sqrt(3)$), the two nonequivalent next-next NN ($t_{3d2}$, $d=a$, along the bonds; and $t_{3d}$, $d=a$ across the hexagon). $t_p$ corresponds to the interplane coupling ($d=c$). Fig.~\ref{figb3} shows these fits as well as the different exchange paths associated with each.

%\subsection{Electronic structure of Ga-Ti-Zn averievite}

The band structure for  Cu$_3$ZnGaTi$_2$O$_{10}$(CsCl) is shown in Fig.~\ref{figb4}. The Fermi level is placed at the Dirac points where a small gap is opened up after inclusion of spin-orbit coupling.

\end{document}